# Collective near-field coupling in infrared-phononic metasurfaces for nano-light canalization

Peining Li[1,2], Guangwei Hu[3,4], Irene Dolado[2], Mykhailo Tymchenko[4], Cheng-Wei Qiu[3], Francisco Javier Alfaro-Mozaz[2], Felix Casanova[2,5], Luis E. Hueso[2,5], Song Liu[6], James H. Edgar[6], Saül Vélez[7], Andrea Alu[4], Rainer Hillenbrand[5,8*]

1 Wuhan National Laboratory for Optoelectronics and School of Optical and Electronic Information, Huazhong University of Science and Technology, Wuhan 430074, China.
2 CIC nanoGUNE BRTA, 20018, Donostia-San Sebastián, Spain.
3 Department of Electrical and Computer Engineering, National University of Singapore, 4 Engineering Drive 3, Singapore 117583, Singapore.
4 Advanced Science Research Center, City University of New York, New York 10031, USA.
5 IKERBASQUE, Basque Foundation for Science, 48013 Bilbao, Spain.
6 Tim Taylor Dept. of Chemical Engineering, Kansas State University, Manhattan, 66506 Kansas, USA.
7 Department of Materials, ETH Zürich, 8093 Zürich, Switzerland
8 CIC NanoGUNE BRTA and UPV/ EHU, 20018, Donostia-San Sebastián, Spain

*Correspondence to: r.hillenbrand@nanogune.eu

**Polaritons – coupled excitations of photons and dipolar matter excitations – can propagate along anisotropic metasurfaces with either hyperbolic or elliptical dispersion. At the transition from hyperbolic to elliptical dispersion (corresponding to a topological transition), various intriguing phenomena are found, such as an enhancement of the photonic density of states, polariton canalization and hyperlensing. Here we investigate theoretically and experimentally the topological transition and the polaritonic coupling of deeply subwavelength elements (structure sizes ~ λ/100, λ being the free-space wavelength) in a uniaxial infrared-phononic metasurface, a grating of hexagonal boron nitride (hBN) nanoribbons. By hyperspectral infrared nanoimaging, we observe, for the first time, a synthetic transverse optical phonon resonance (that is, the strong collective near-field coupling of the nanoribbons) in the middle of the hBN Reststrahlen band, yielding a topological transition from hyperbolic to elliptical dispersion. We further visualize and characterize the spatial evolution of a deeply subwavelength canalization mode (a spatial confinement of about 300 nm, ~ λ/22) near the transition frequency, which is a collimated polariton that is the basis for hyperlensing and diffraction-less propagation. Our results provide fundamental insights into the role of polaritonic near-field coupling in metasurfaces for creating topological transitions and polariton canalization.**

## Introduction

Uniaxial metasurfaces are thin layers of engineered subwavelength structures, whose in-plane effective permittivity tensor components are unequal ($\varepsilon_{eff,x} \neq \varepsilon_{eff,y}$), and thus support different types of in-plane anisotropic polaritons ($1$–$10$). When both $\varepsilon_{eff,x}$ and $\varepsilon_{eff,y}$ are negative but with different absolute values, the polaritons propagating along the metasurface exhibit an elliptically-shaped dispersion diagram, i.e., the polariton momentum $\boldsymbol{k}$ describes an elliptical isofrequency contour (IFC) in $k$-space ($6$, $9$, $11$). On the other hand, when $\varepsilon_{eff,x}$ and $\varepsilon_{eff,y}$ have opposite signs, polaritons possess a so-called hyperbolic dispersion ($\boldsymbol{k}$ describes hyperbolic IFCs in $k$-space) ($1$, $4$–$10$, $12$), exhibiting increased polariton confinement and ray-like anisotropic propagation along the surface. These two types of anisotropic metasurfaces can be applied, for example, to enhance optical birefringence ($13$, $14$), to control light polarization ($15$), for nanoscale directional polariton guiding ($5$, $8$, $9$, $16$, $17$), and for subwavelength-scale optical imaging ($9$, $18$).

A particularly interesting regime arises when uniaxial metasurfaces exhibit a topological transition of the IFCs upon frequency variation ($8$, $9$, $19$), changing from hyperbolic to elliptical. It offers unique opportunities in nanophotonics, for example, for enhancing the local photonic density of states ($8$, $9$, $19$) and for supporting deeply subwavelength canalization modes ($16$, $20$). These canalization modes can exhibit extremely anisotropic in-plane polariton momenta, which results in nanoscale and nearly diffraction-free electromagnetic energy transport with applications in hyperlensing ($8$, $9$, $16$) and control of near-field heat transfer ($9$). It has been shown theoretically that the topological transition and the canalization modes are determined by the polaritonic near-field coupling of subwavelength elements comprising the metasurfaces ($7$, $9$, $16$). However, so far, the polaritonic coupling governing the topological transition has not been experimentally demonstrated. Only few experimental studies at microwave frequencies have visualized weakly confined canalization modes at 10 GHz ($10$). Here, we demonstrate that the strong collective near-field coupling of subwavelength elements in an infrared-phononic metasurface (a hBN nanograting) yields a synthetic optical phonon resonance and subsequently a topological transition. By hyperspectral nanoimaging, we are able to observe the topological transition and strong coupling of nanoribbons (the metasurface elements in our case) in spatial and spectral domains. In particular, we provide the first real-space optical images of deeply subwavelength canalization polaritons, which thus experimentally demonstrate that these modes are the consequence of strong collective near-field polaritonic coupling and the associated strong nonlocal response in metasurface.

## Results

Boron nitride exhibits a negative and isotropic in-plane permittivity ($\varepsilon_{hBN} = \varepsilon_x = \varepsilon_y < 0$) in its upper mid-infrared Reststrahlen band (the frequency region between transversal and longitudinal optical phonon frequencies, TO and LO, respectively), where phonon polaritons (PhPs) exist ($5$, $21$–$27$). Patterning a thin hBN flake into a periodic array of nanoribbons (nanograting) creates an infrared metasurface with strong in-plan anisotropy ($5$) (Fig. 1A, ribbon width $w$=70nm, gap size $g$=30nm, thickness $h$=20nm). By considering the polaritonic near-field coupling of the ribbons and the strong nonlocal effect induced by the periodic structuring (the effective permittivity depends on the polariton momentum $\boldsymbol{k}$ that is controlled by the grating period $L$, see discussions in ref. ($8$, $9$)), the effective in-plane anisotropic permittivity of this metasurface can be described by a modified effective medium model ($8$, $9$)

$$\varepsilon_{\text{eff,x}} = \left(\frac{1-\xi}{\varepsilon_{\text{hBN}}/\varepsilon_c} + \frac{\xi}{\varepsilon_{\text{air}}}\right)^{-1}, \qquad \varepsilon_{\text{eff,y}} = (1-\xi)\varepsilon_{\text{hBN}} + \xi\varepsilon_{\text{air}} \qquad (1)$$

where $\xi = \frac{g}{L}$ is the filling factor. $\varepsilon_c$ is a coefficient capturing the near-field coupling of neighboring ribbons and the nonlocal effect associated with the grating period, $\varepsilon_c = \frac{2L}{\pi h}\log\left[\csc\left(\frac{\pi\xi}{2}\right)\right]$. Note that the hBN grating can be well modeled as a uniaxial sheet owing to the very small flake thickness (28).

The analytically calculated effective permittivity $\varepsilon_{\text{eff,x}}$ (red line in Fig.1B) reveals the emergence of a new TO phonon frequency at $\omega_{\text{STO}} = 1478$ cm$^{-1}$, perpendicular to the ribbons. It results from the strong collective near-field coupling of the dipolar PhP resonance of the individual hBN ribbons (figs. S1 to S3), analog to the TO phonon in polar crystals. For this reason, we name this collective mode a synthetic TO phonon (STO) (29). Note that a standard effective medium model based on Maxwell-Garnett (MG) approximation also predicts the existence of the STO resonance (dashed grey line in Fig.1B). However, it is shifted by about 50 cm$^{-1}$ to higher frequencies because polaritonic near-field coupling and nonlocal effects induced by the grating geometry are not considered. As shown below, the STO predicted by our modified effective medium model (Equation 1) is in an excellent agreement with both numerical and experimental results (See Fig.1C, Fig.3, and fig. S1 and S4).

As a result of the STO, the photonic local density of states (PLDOS) on the metasurface differs dramatically from the one of natural hBN flakes, as confirmed by the calculated PLDOS spectra shown in Fig.1C. In the simulations a point dipole source is placed at the height of 200 nm above the surface (see Materials and Methods). The PLDOS spectrum of the hBN layer exhibits a strong peak around $\omega = 1450$ cm$^{-1}$ (dashed grey line), due to the excitation of a fundamental PhP "waveguide" mode in the hBN slab (21, 25). In contrast, the metasurface (modeled as an effective medium according to Equation 1) exhibits two PLDOS peaks located on either side of the STO (at 1430 and 1480 cm$^{-1}$, blue line). These two peaks are verified by a full-wave numerical simulation using a real grating structure (red line) and indicate that two distinct PhP modes are excited on the metasurface (fig.S3). This result further corroborates the validity of the modified effective medium theory described by Eq. 1 (in contrast to the standard MG theory, which fails in quantitative prediction of the STO frequency).

Below the STO, the dipole excites PhPs possessing an in-plane hyperbolic dispersion (5) (because Re($\varepsilon_{\text{eff,x}}$) > 0 and Re($\varepsilon_{\text{eff,y}}$) < 0), which are formed by near-field coupling of the waveguide polaritons propagating along individual nanoribbons (note that this coupling is weak and thus yields only a positive value for $\varepsilon_{\text{eff,x}}$). The propagation of these in-plane hyperbolic PhPs (HPhPs) is highly anisotropic along the metasurface, exhibiting the typical ray pattern of hyperbolic polaritons (Fig.1D). Fourier transform (FT) of the near-field distribution $E_z$ indeed yields a hyperbolic IFC describing the polariton momentum $\boldsymbol{k}$ (Fig.1E) in momentum space at fixed frequency.

Above the STO, the dipole-excited PhPs have extremely elliptical in-plane dispersion, owing to $\varepsilon_{\text{eff,x}}$ and $\varepsilon_{\text{eff,y}}$ being negative but with largely different absolute values. The elliptical PhPs (EPhPs) are due to the strong collective near-field coupling (yielding a negative $\varepsilon_{\text{eff,x}}$) of individual nanoresonators (the nanoribbons exhibiting the Fabry–Pérot polariton resonances, see our verifications in Fig.3 and fig.S7), which are visualized by simulation in Fig.1F, where the dipole source launches a collimated polariton beam with lateral confinement of about 150 nm ($\sim\lambda/45$). The FT of the near-field distribution confirms the polaritons´ extreme elliptical IFC in $\boldsymbol{k}$-space (Fig. 1G; note that the IFC is not a perfectly closed ellipse owing to polariton damping by intrinsic material loss).

Although it may be expected that EPhPs near the STO may suffer from the large imaginary part of $\varepsilon_{\mathrm{eff},x}$ (fig. S5), their absolute propagation length is comparable to the one of HPhPs (Fig.1D and fig.S6). This can be explained by the large negative real part of $\varepsilon_{\mathrm{eff},x}$ near the STO, which actually reduces the field confinement inside the material, repelling the fields and hence reducing the absorption (16). By increasing the frequency, the EPhPs become more confined and decay faster (Fig. 1H, more simulations shown in fig. S4), exhibiting a weaker ellipticity (Fig. 1I) owing to the decreasing figure of merit $|\mathrm{Re}(\varepsilon_{\mathrm{eff},x})|/|\mathrm{Re}(\varepsilon_{\mathrm{eff},y})|$. The highly collimated EPhP modes described in Figs. 1F,H are also known as canalization modes (16, 20). They have been explored in bulk metamaterials (20, 30) or two-dimensional metasurfaces (16) for various applications, including hyperlensing (9, 16, 31) and subwavelength focusing (30). Theory also has been predicting the canalization of plasmon polaritons on metallic metasurfaces (graphene (16), black phosphorus (12), or metals (7)) at optical (visible and infrared) frequencies, which, however, has not been experimentally demonstrated. In our work we do not only theoretically predict the canalization of *low-loss deeply-confined phonon polaritons* but also demonstrate them experimentally via infrared nanoimaging.

To image the canalization of EPhP modes, we fabricated a hBN-metasurface ($w$=75nm and $g$=25nm) on a 20-nm-thick flake of monoisotopic low-loss hBN (5, 27, 32, 33) (schematics are shown in Fig.2A; for details see Materials and Methods). We first performed polariton-interferometry on the metasurface with a scattering-type scanning near-field optical microscope (s-SNOM) (34, 35). The metallized tip of an atomi   jjfjfjfhggjgjhggjgjhjgjhgjgc force microscope (AFM) is illuminated by a *p*-polarized infrared laser beam, operating as an infrared nanoantenna that concentrates the incident field at its sharp apex, yielding a nanoscale near-field spot for launching the polaritons. The polaritons propagate away and are reflected at the boundaries of the metasurface. They propagate back and interfere with the polariton field below the tip, forming interference fringes (with spacing equals to half the polariton wavelength, $\lambda_p/2$), which are visualized by recording the tip-scattered field as a function of tip position (34, 35).

Figure 2B and C present polariton-interferometry images (amplitude signals, *s*) taken at frequencies within the HPhP and EPhP regions, respectively at 1415 and 1510 cm$^{-1}$. On the grating area, we observe only horizontal fringes in the HPhP spectral range (Fig. 2B), and only vertical fringes in the EPhP spectral range (Fig. 2C). The two distinct fringe orientations reveal the different propagation directions of the polaritons, as predicted in our simulations shown in Fig. 1D, F, and H. They provide a first experimental indication for the existence of EPhPs above STO, and of the topological transition in momentum space.

To explore the frequency range and dispersion of the two types of polaritons, we recorded near-field spectroscopic line scans (along the lines marked in Fig.2B and C). In the line scan parallel to the ribbons (Fig.3A), we observe a horizontal feature round $\omega = 1400$ cm$^{-1}$, matching well the TO phonon of hBN (see Fig.3I for a comparison of experimental and simulated near-field spectra). Above the TO, we observe a series of fringes (indicated by dashed black curves), whose spacing is reducing with increasing frequencies. They reveal the in-plane HPhPs propagating parallel to the ribbons, whose wavelength is shrinking at higher frequencies. Around $\omega = 1500$ cm$^{-1}$ we observe a broad horizontal (non-dispersive) feature that fits well the STO (see Fig.3I).

In the line scan perpendicular to the grating (Fig.3B), we again observe the horizontal feature corresponding to the TO phonon. In the whole spectral region between TO and STO (the HPhP range), we do not observe fringes, indicating the absence of polariton propagation perpendicular to the grating. As in Fig.3A, we observe the horizontal feature at the STO (see also fig. S7). More interestingly, we see an interference fringe (marked by a blue dashed curve) above the STO. Its distance to the boundary decreases with increasing frequency, corroborating that the fringes in Fig. 2C indeed reveal a polariton propagating perpendicular to the grating. We note that the fringe intensity is modulated by the ribbons. Inside the gaps between ribbons, the fringe intensity is higher, as here the tip more efficiently launches polariton propagating perpendicular to the ribbons. The tip-launched polariton is reflected at boundary, giving rise to the observed fringe (illustration in Fig. 3H). This polariton mode can be actually understood as induced by energy tunneling via near-field polaritonic coupling of neighboring ribbons, similar to energy transport in plasmonic particle chains(*36*). We further observe a horizontal series of bright spots at $\omega = 1570$ cm$^{-1}$. However, this feature is not accompanied by interference fringes at higher frequencies, indicating a purely localized mode. A zoom-in image and analysis (fig. S7) show that the bright spots reveal a localized second-order transverse polariton mode of the ribbons, as illustrated in Fig. 3G.

Numerical simulations of spectroscopic line scans (Fig.3C,D; a dipole source was scanned above the grating and the field below the dipole was recorded, see Methods) confirm our experimental results, particularly the interference fringe (dashed blue curve) above the STO, and the localized ribbon resonance around $\omega = 1570$ cm$^{-1}$ (marked by red arrow). We repeated the simulations for a metasurface described by our effective medium theory (Fig.3E, F), reproducing well the results of Fig.3C and D, but avoiding the signal modulation introduced by the grating features, thus yielding a clearer map. The good agreement confirms the validity of our nonlocal effective medium model, especially important to properly capture the properties of canalization polaritons near the topological transition.

Altogether, Fig.3A and B experimentally verify two spectral regions (separated by the STO) within the hBN Reststrahlen band, in which two different types of polaritons exist. Their different propagation directions provide experimental evidence that the IFC of the polariton momentum undergo a topological transition across the STO.

To experimentally visualize the canalization mode near the STO, predicted in Fig.1 F and H, we imaged the polariton emitted from an infrared antenna (a gold rod) on another metasurface (fabricated together with the one in Fig. 2 on the same flake, topography in Fig.4A). The antenna concentrates the mid-infrared illumination to nanoscale spots at its antenna extremities, acting as a nanoscale source for launching the polaritons. Figure 4B-D presents the experimental images of the antenna-launched polaritons propagating and decaying along the metasurface. In the grating area, periodic horizontal bright lines are observed. As explained in Fig.3, they correspond to the strong near-fields inside the gaps, because of enhanced tip-polariton coupling. More importantly, we indeed observe the deep-subwavelength canalization EPhP — a collimated polariton beam (with a lateral confinement of 310 nm, $\sim\lambda/22$, see Fig.4J; see also the FT results in fig.S8) emitted from the antenna extremity. At $\omega = 1495$ cm$^{-1}$ it is collimated over at least five ribbons (Fig. 4B). At higher frequencies, the polaritons are more confined and thus decay faster (Fig.4C,D), but they still extend farther than the antenna fields decaying on a dielectric substrate ($\varepsilon_{hBN} \approx 1$ at $\omega = 1735$ cm$^{-1}$, Fig.4H). In a control experiment, we imaged the polaritons launched by the antenna on the unpatterned hBN (Fig.4G, topography in Fig.4F), showing radial propagation along the hBN, in striking contrast to the canalization modes (Fig.4B-D).

We numerically verified the antenna-launched canalization polaritons by simulating the electromagnetic near-field distribution around the antenna on the grating structure (Fig.4E). On the other hand, the simulated canalization mode exhibits a longer decay length than that experimentally realized. This can be explained by higher damping in the experiment caused by fabrication uncertainties and material damage from etching.

Another important insight of our experiments (Fig.4B-D) is the direct visualization of energy flow transported along a chain of coupled polaritonic nanoresonators(*36*). More precisely, we use the antenna to locally illuminate the first ribbon. Energy flows directionally to the next ribbons owing to the polaritonic near-field coupling and the extreme in-plane anisotropy of the canalization mode, which avoids energy spreading in other directions. The electric-field decay length for this process is quantified to be about 220 nm (blue line in Fig.4I, background subtracted, see fig.S9) by fitting the near-field profile (along the vertical dashed blue line in Fig.4B) with an exponential decay (dashed red line in Fig.4I). This value is much larger than the one of antenna fields on the bare dielectric substrate (green line in Fig.4I, decay length < 50 nm). Our results therefore provide the first real-space observation of energy flow through coupled phonon-polaritonic nanoresonators at infrared frequencies, with important consequences for the development of infrared photonic and thermal devices based on near-field polaritonic coupling. These results also confirm the important role of strong coupling between neighboring resonators to achieve extreme anisotropy and canalization. The consequent nonlocality, well captured by our homogenized metasurface model, plays an important role in the physics demonstrated in this paper.

**Discussion**

STOs (strong collective coupling of the metasurface elements) and topological transitions may also be envisioned in other types of metasurfaces, for instance based on strongly coupled graphene (or black phosphorus) nanoresonators (*8, 9*), which may lead to electrically-tunable collective resonances and canalization polaritons for sensing and thermal emission applications at infrared and THz frequencies. The demonstrated deep-subwavelength canalization polaritons hold promise for many exciting applications, including in-plane hyperlensing (*8, 9*), on-chip collimated polariton emitting, waveguiding and focusing (*8, 9, 16*).

## Materials and methods

### Nanoimaging

We used a commercial s-SNOM system (from Neaspec GmbH) based on an atomic force microscope (AFM). The Pt-coated AFM tip (oscillating vertically at a frequency $\Omega \approx 270\ kHz$) was illuminated by light from a wavelength-tunable continuous-wave quantum cascade laser. The backscattered light was collected with a pseudo-heterodyne interferometer (37). To suppress background contribution in the tip-scattered field, the interferometric detector signal was demodulated at a higher harmonic $n\Omega$ ($n \geq 2$), yielding near-field amplitude $s_n$ and phase $\varphi_n$ images. Fig. 2 and 4 show amplitude $s_3$ images.

### Spatio-spectral near-field observation

For spatio-spectral observation shown in Fig.3, the s-SNOM tip and sample were illuminated with a broadband mid-infrared laser. The tip-scattered signal was analyzed with an asymmetric Fourier transform spectrometer (based on a Michelson interferometer), in which tip and sample were located in one of the interferometer arms (24, 25). An interferogram was measured by recording the demodulated detector signal (the harmonic $3\Omega$ for background suppression) as a function of the position of the reference mirror, at a fixed tip position. Subsequent Fourier transform of the recorded interferogram yields a complex-valued near-field point spectrum (24, 25). We scanned the tip parallel or perpendicular to the hBN ribbons, respectively. At each tip position, we recorded a complex-valued near-field point spectrum. By plotting the recorded near-field amplitude $s_3$ as a function of the tip position and the operation frequency, we obtained the images shown in Fig. 3a and b.

### Sample preparation

For experiments we used isotopically ($^{10}$B) enriched h-BN (details of the growing process can be found in ref. (32) and (33)), which exhibits ultra-low-loss phonon polaritons (27). We fabricated infrared metasurfaces by the etching process reported in ref. (5).

### Numerical simulations

We used a finite-elements method based software (COMSOL Multiphysics) for simulations. In the simulations, the permittivity of the isotopically enriched h-BN was taken from ref. (5). The details of the simulations are provided in note S1.

### Acknowledgements

The authors acknowledge financial support from the Spanish Ministry of Science, Innovation and Universities (national projects MAT2017-88358-C3, MAT2015-65159-R, MAT2015-65525-R, RTI2018-094830-B-100, RTI2018-094861-B-100, and the project MDM-2016-0618 of the Marie de Maeztu Units of Excellence Program) and the Basque Government (PhD fellowship PRE 2018 2 0253). Further, support from the Materials Engineering and Processing program of the National Science Foundation, award number CMMI 1538127, and the II−VI Foundation is greatly appreciated.

### Competing interests

R.H. is co-founder of Neaspec GmbH, a company producing scattering-type scanning near-field optical microscope systems, such as the one used in this study. The remaining authors declare no competing interests.

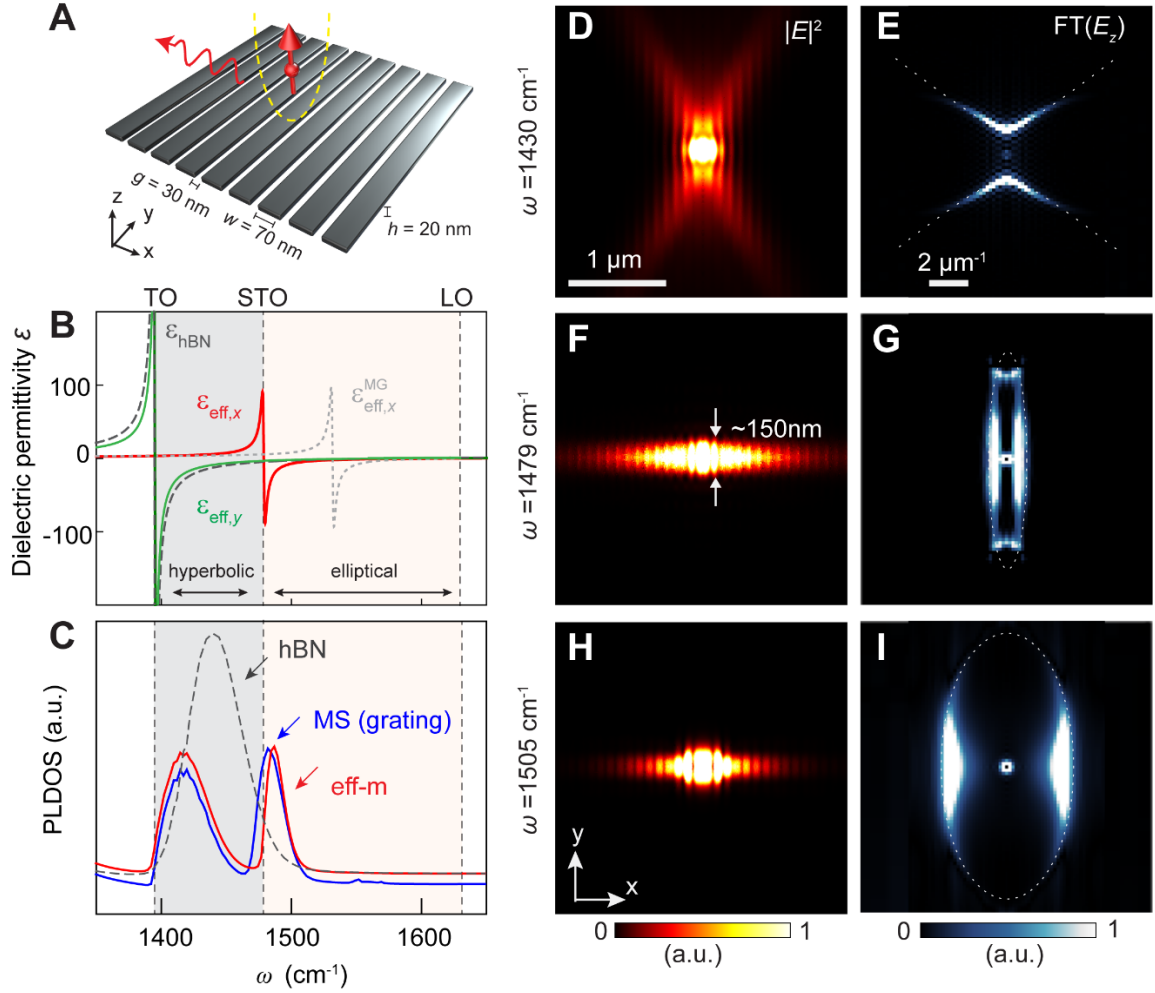

**Figure 1 Polariton topological transition and canalization in hBN metasurfaces.** (**A**) Schematic of a 20-nm-thick hBN metasurface based on grating nanostructures. (**B**) Anisotropic effective dielectric permittivities (real parts) of the metasurface calculated using Equation 1, $\varepsilon_{\text{eff},x}$ (red line) and $\varepsilon_{\text{eff},y}$ (green). Permittivity of unpatterned hBN ($\varepsilon_{\text{hBN}}$, dashed brown line) and effective permittivity of the metasurface based on Maxwell-Garnett approximation ($\varepsilon_{eff}^{MG}$, dashed gray line) are provided for comparison. (**C**) Simulated PLDOS on the grating metasurface (MS), the effective medium (eff-m) and the unpatterned hBN, respectively. (**D**), (**F**) and (**H**) Near-field intensity distribution of dipole-launched polaritons at three different frequencies. (**E**), (**G**), and (**H**) Absolute value of the Fourier transform (FT) of the simulated near-field distribution $E_z$ (see fig. S4). Dotted hyperbola and ellipses are guides to the eye.

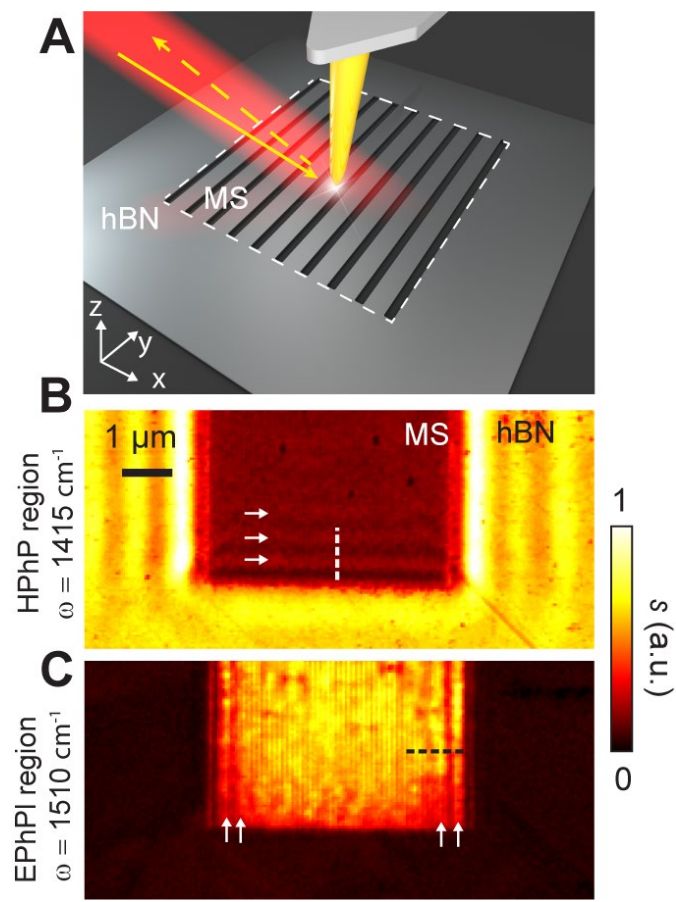

**Figure 2 Near-field imaging of polariton evolution in a hBN metasurface. (A)** Schematic of the near-field nanoimaging experiment. **(B** and **C)** Near-field images measured at two different frequencies, $\omega$=1415 cm$^{-1}$ (HPhP region) and $\omega$=1510 cm$^{-1}$ (EPhP region). White arrows indicate the polariton fringes observed on the metasurface.

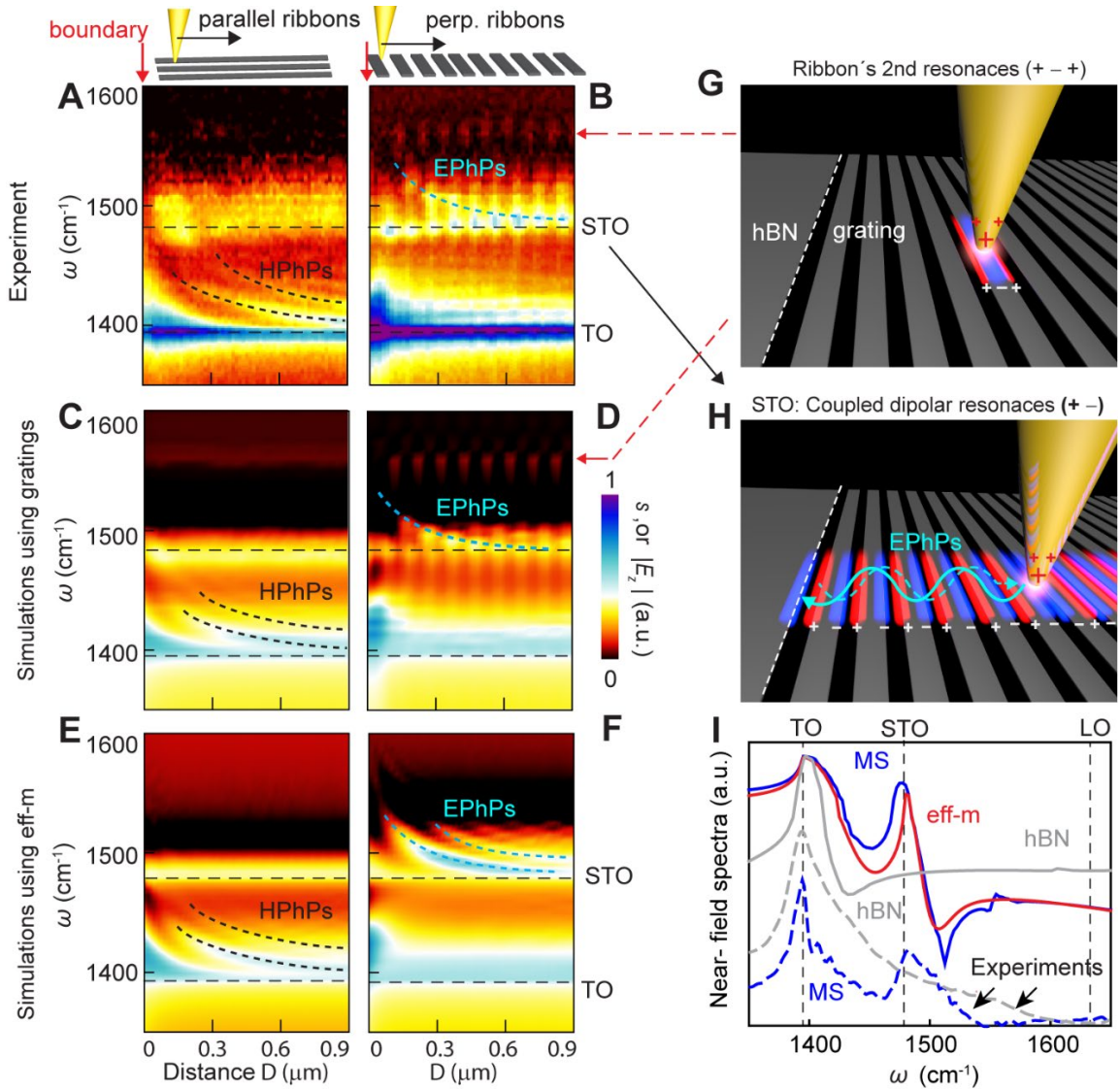

**Figure 3 Hyperspectral nanoimaging of polariton evolution in a hBN metasurface.** Near-field spectroscopic line scans taken parallel (**A**) and perpendicular (**B**) to the ribbons, as indicated in the top schematics. (**C** to **F**), Simulated line scans based on two different models, **c**) and **d**) for real grating structures, (**E** and **F**) for the effective medium (eff-m). (**G** and **H**) are schematics of tip-excitation of the second order and dipolar PhP resonances of the hBN ribbons, respectively. (**I**) Bottom dashed lines, experimental near-field spectra taken from the metasurface (on a gap of the grating, dashed blue line) and on the unstructured hBN (dashed grey line). Top solid lines, simulated near-field spectra of the grating (blue line), of eff-m (red line) and of the unstructured hBN (grey line).

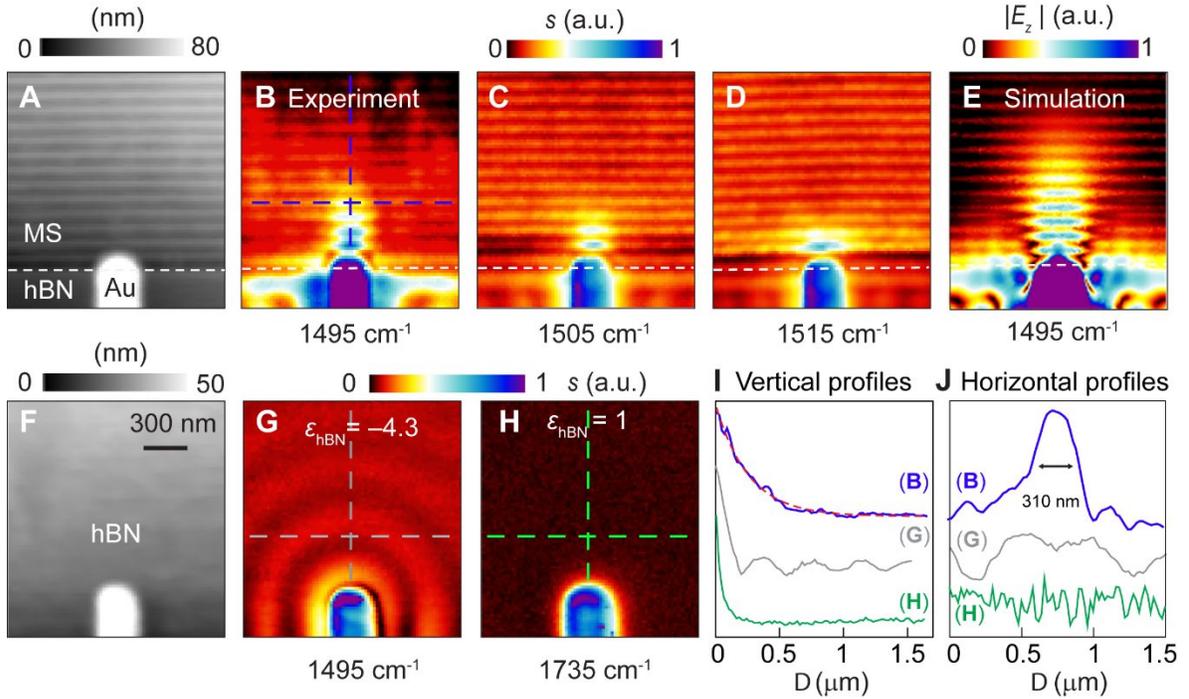

**Figure 4** **Real-space imaging of polariton canalization in a hBN metasurface.** (**A**) Topography of the sample. (**B** to **D**) Experimental near-field distribution of antenna-launched elliptical polaritons on the metasurface at three different frequencies. White dashed lines mark the boundary of the metasurface. (**E**) Simulated near-field distribution of the antenna-launched canalization EPhPs on the metasurface. (**F**), (**G**) and (**H**) Topography and experimental near-field images (at two different frequencies) for the case of an antenna located on an unpatterned area of the same hBN flake. (**I**) Blue line is the profile taken from the image shown in the panel (**B**) (along the vertical dashed blue line), which can be fitted by an expontial decay (dashed red line). Grey line, the profile taken from the panel (**G**) (along the vertical dashed grey line). Green line, the profile taken from the panel (**H**) (along the vertical dashed green line). (**J**) Horizontal line profiles taken from the panels (**B**), (**G**) and (**H**), repectively.

Supplementary Materials for

# Collective near-field coupling in infrared-phononic metasurfaces for nano-light canalization

*Peining Li[1,2], Guangwei Hu[3,4], Irene Dolado[2], Mykhailo Tymchenko[4], Cheng-Wei Qiu[3], Francisco Javier Alfaro-Mozaz[2], Felix Casanova[2,5], Luis E. Hueso[2,5], Song Liu[6], James H. Edgar[6], Saül Vélez[7], Andrea Alù[4], Rainer Hillenbrand[5,8*]*

1 School of Optical and Electronic Information, Huazhong University of Science and Technology, 430074 Wuhan, China.
2 CIC nanoGUNE BRTA, 20018, Donostia-San Sebastián, Spain.
3 Department of Electrical and Computer Engineering, National University of Singapore, 4 Engineering Drive 3, Singapore 117583, Singapore.
4 Advanced Science Research Center, City University of New York, New York 10031, USA.
5 IKERBASQUE, Basque Foundation for Science, 48013 Bilbao, Spain.
6 Tim Taylor Dept. of Chemical Engineering, Kansas State University, Manhattan, 66506 Kansas, USA.
7 Department of Materials, ETH Zürich, 8093 Zürich, Switzerland
8 CIC NanoGUNE BRTA and UPV/ EHU, 20018, Donostia-San Sebastián, Spain

*Correspondence to: r.hillenbrand@nanogune.eu

**Note S1. Simulation details**

For simulations in Fig. 1, an electric dipole source at the height of 200 nm above the surface is used to excite the polaritons. The metasurface is modeled as either real grating nanostructures or an effective medium (see descriptions in the corresponding text). The near-field distribution is taken from the plane at the height of 50 nm above the surface. For calculating the PLDOS spectra shown in Fig.1C, we use the method introduced by Peragut, F. *et al.* (*Optica* 4, 1558 (2017)). Because of the $z$-axis orientation of the dipole source, we investigated the $z$ component of the PLDOS that is proportional to the $z$ component of the Green tensor (Im[G($r,r,\omega$)]). The quantity Im[$G_{zz}(r,r,\omega)$] is also proportional to the $z$ component of electric field (the real part, Re[$E_z$]) at the position of the dipole source. Thus, by normalizing the numerically simulated electric field Re[$E_z$] at the position of the dipole source to that obtained in vacuum, we obtain the PLDOS normalized by the PLDOS in vacuum as shown in Fig.1C of the main text.

For simulations in Fig. 3, we scanned the dipole (200 nm above the surface) over the metasurface (either the grating or the effective medium) along two different directions, respectively. The electric field (at the height of 50 nm above the surface) below the dipole is recorded as a function of the spatial position and the operation frequency, yielding the figures shown in Fig.3C to F.

For simulations in Fig. 4E, we used an Au antenna on the grating to excite the canalization mode. For modeling the grating metasurface, we used the ribbon width $w = 60$ nm and the gap size $g = 40$ nm according to the fabricated structures. We used a plane wave to illuminate the antenna and the metasurface. The near-field distribution is taken from the plane 50 nm above the surface. All simulations took into account the Si/SiO$_2$ (250 nm thick) substrate.

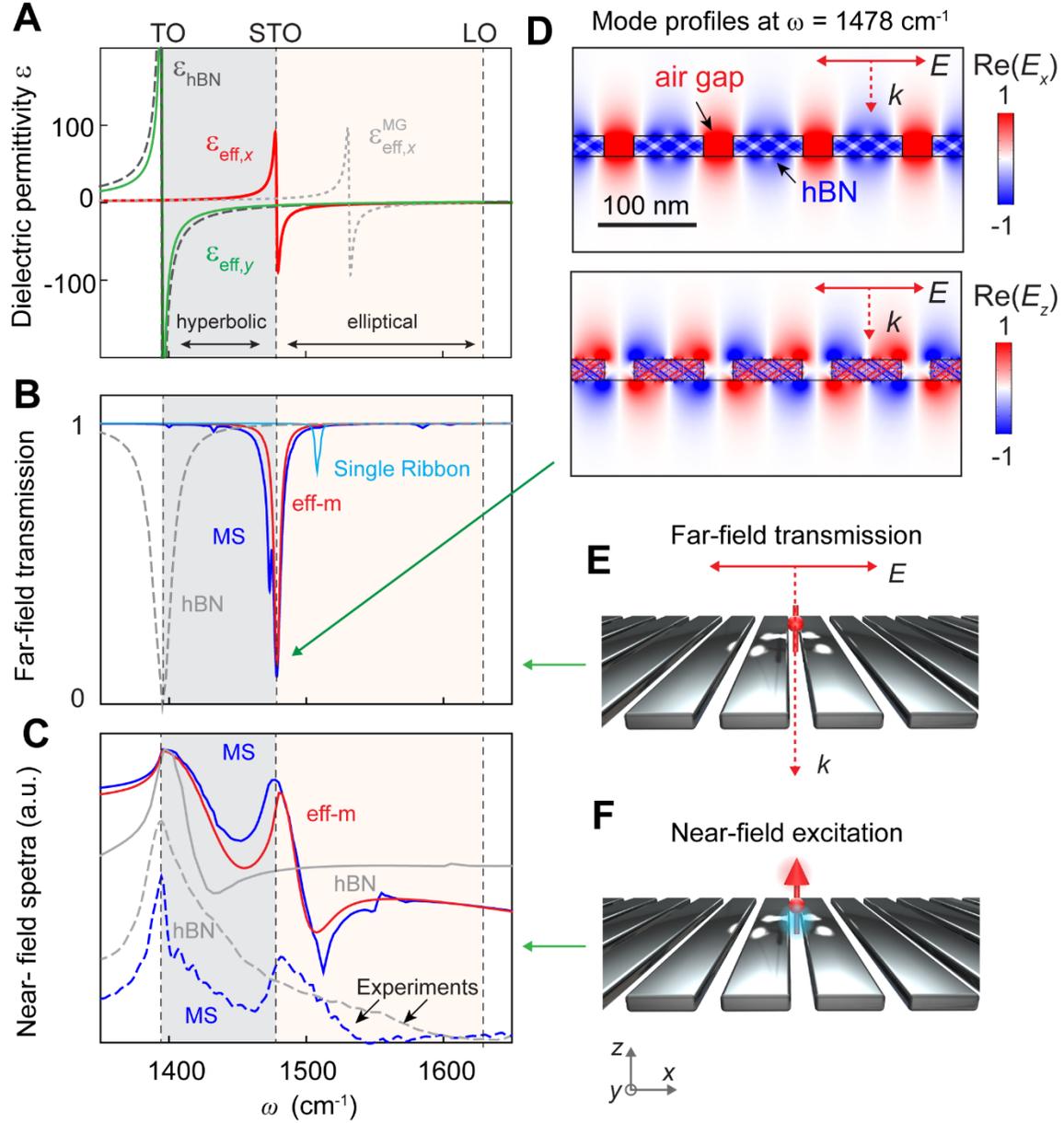

**Figure S1. Comparison of far-field and near-field spectra and effective permittivity. (A)** Elements of the permittivity tensor as shown in Fig.1b of the main text. **(B)** Simulated far-field transmission spectra of the grating metasurface (blue), the metasurface modeled as an effective medium (eff-m, red), the un-patterned hBN (dashed grey) and single hBN nanoribbon (light blue). In these four cases, the layer thickness is the same as 20 nm. **(C)** Experiemental and simulated near-field spectra as presented in Fig.3I of the main text. **(D)** Simulated near-field distribution (the real parts of $E_x$ and $E_z$) of the nanograting excited by a plane wave illumination at the frequency of the STO resonance $\omega$ =1478 cm⁻¹. **(E and F)** Schematics of the far-field transmission through the grating and near-field dipole excitation of the grating's resonance, respectively.

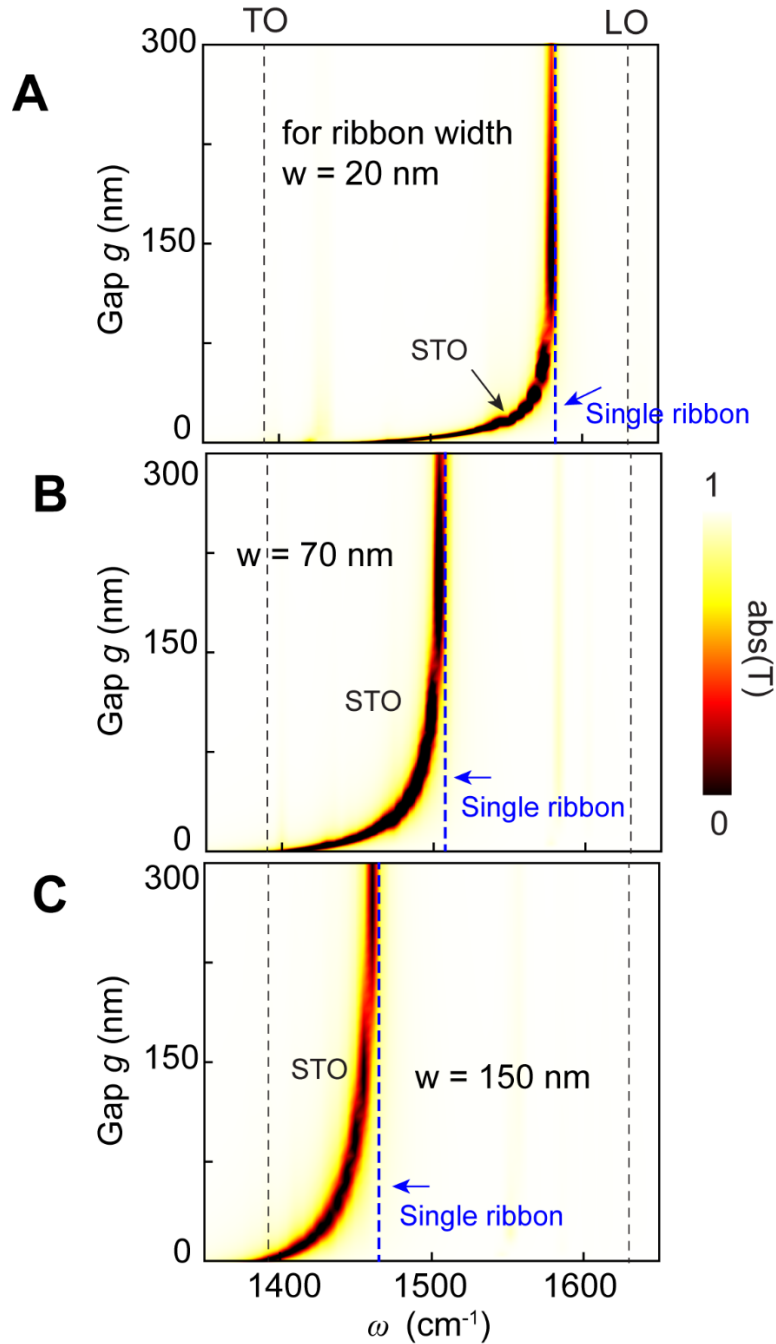

**Figure S2. Tuning the STO resonance with the gap size and the ribbon width**. (**A** to **C**) Simulated far-field transmittance of the grating metasurface as a function of the air-gap size and the frequency, for (**A**) ribbon width $w$=20 nm, (**B**) $w$=70 nm, (**C**) $w$=150 nm. The vertical blue lines mark the resonance positions of single nanoribbons. This figure shows that the STO resonance of the metasurface can be well tuned by adjusting the sizes of both the gap and the ribbon width. For a fixed ribbon width, the resonance of single ribbon determines the upper limit of tuning the STO resonance with the gap size.

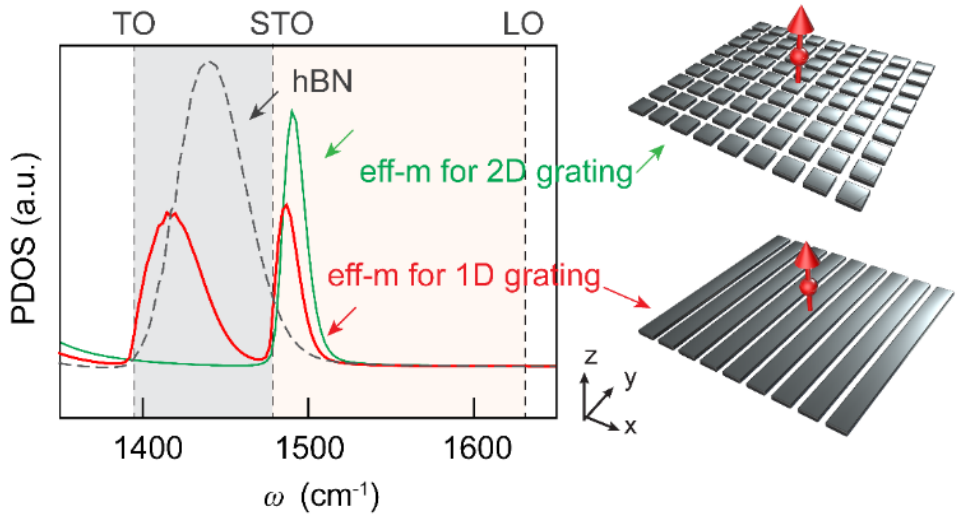

**Figure S3. Simulated PDOS spectra of 1D and 2D grating.** The 1D grating exhibits two PDOS peaks that are due to the two different PhP modes (HPhPs and EPhPs) excited on the metasurface (modeled as an effective medium). In contrast, the 2D grating only exhibits a single PLDOS. This is because the 2D grating is the in-plane symmetrical, which thus supports an in-plane isotropic PhP mode. Therefore, this figure shows that the metasurface has great possibilities to engineer the polaritons propagating along the surface.

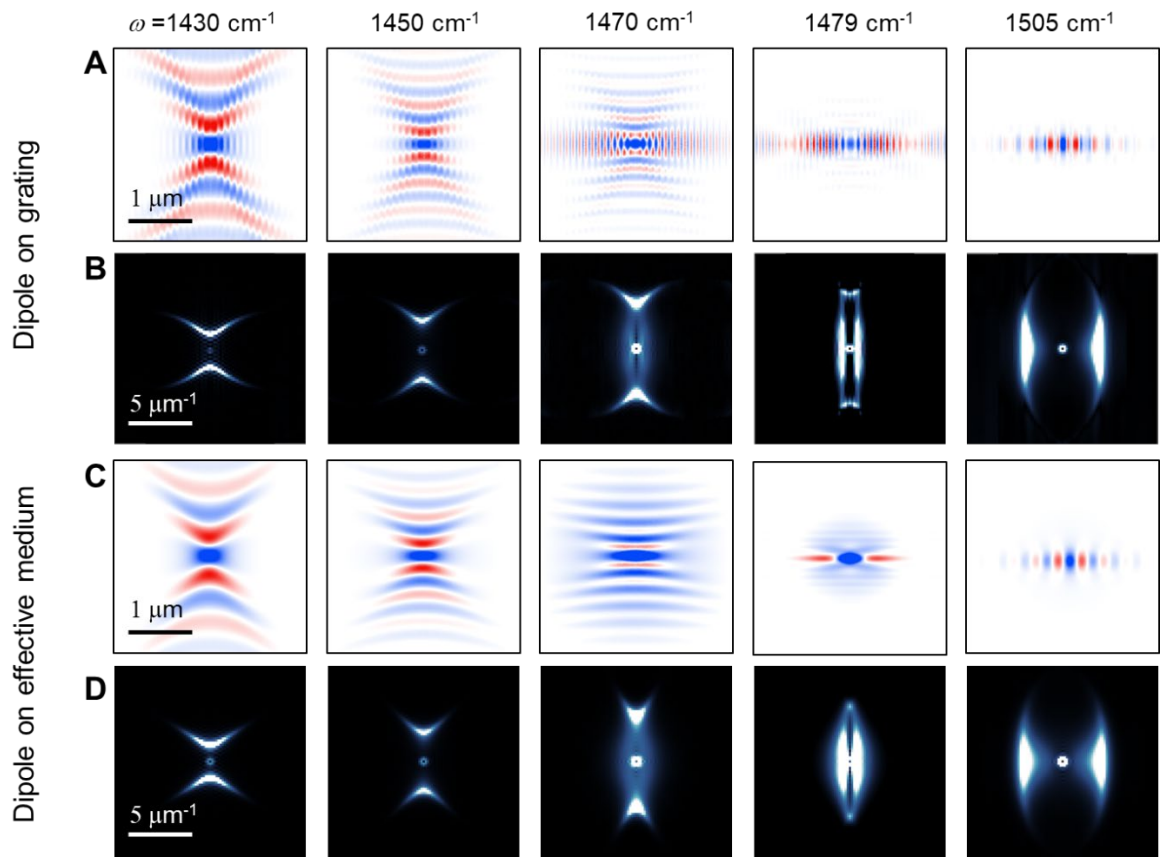

**Figure S4. Simulated near-field spatial evolution of PhPs launched by a dipole source:** on (**A**) the grating and on (**C**) the effective medium (calculated by Equation 1 in the main text). (**B**) and (**D**), the Fourier transform of the (**A**) and (**C**), respectively.

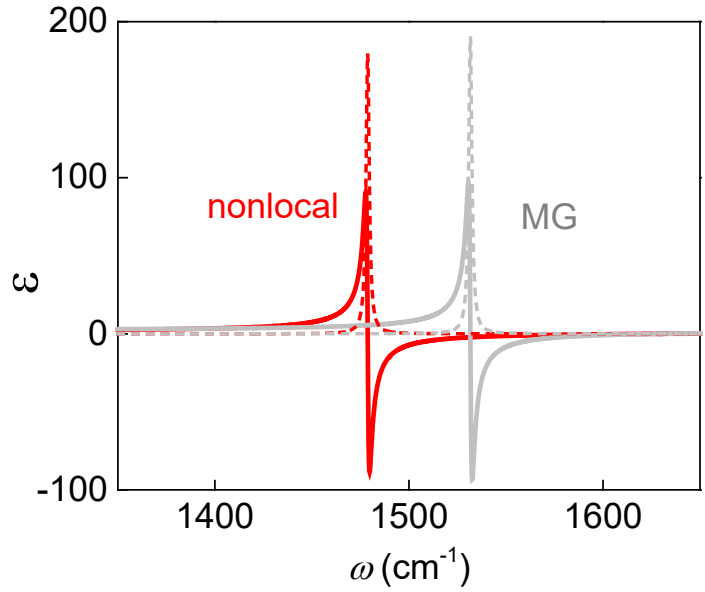

**Figure S5. Calculated effective permittivity $\varepsilon_{\text{eff},x}$ based on two distinct models.** Nonlocal effective medium model, Equation 1 (red) and standard effective medium model based on Maxwell-Garnett (MG) approximation (grey). Solid lines, real parts; dashed lines, imaginary parts. Both models predict the existence of STO resonance. However, the one predicted by the MG model is shifted by at least 50 cm⁻¹ to a wrong spectral position (as we experimentally corroborated in Fig.3 of the main text), because it does not consider the polaritonic coupling of the ribbons and the nonlocal effects induced by the structuring.

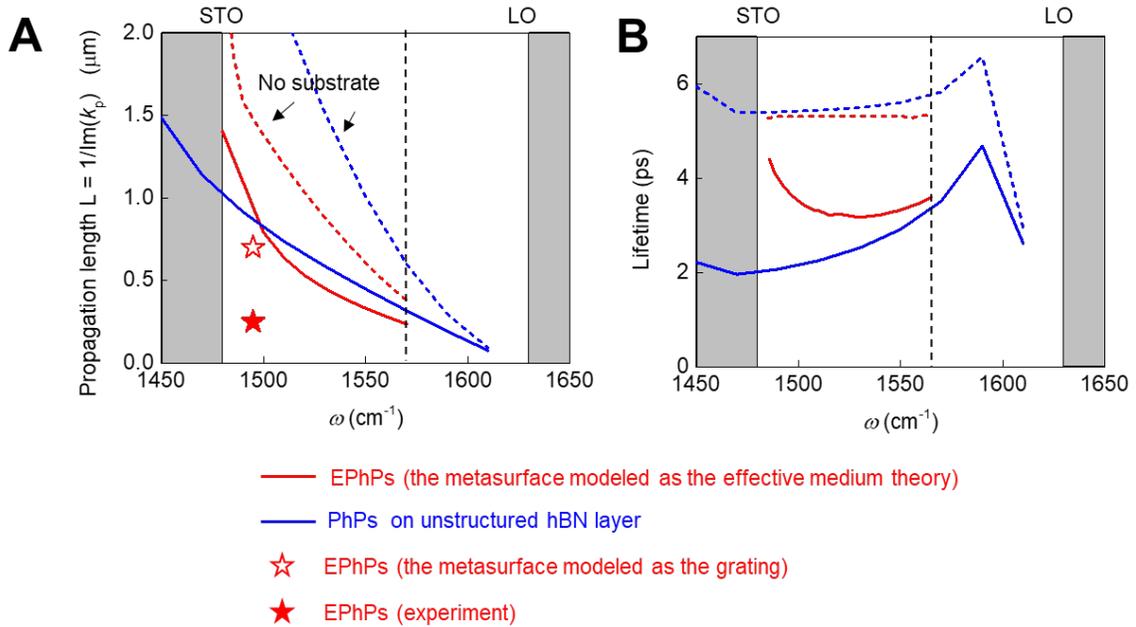

**Figure S6. Comparison of propagation length (A) and lifetime (B) of phonon-polariton on the metasurface and unstructured slab.** We first calculate the propagation length and lifetime of EPhPs using the COMSOL software with the effective permittivity calculated from our nonlocal model (vertical dashed lines in the figures mark the frequency limit that our model is valid**).** In the case with the SiO$_2$/Si substrate, the EPhPs (red solid line) on the metasurface exhibit a similar propagation length and a longer lifetime compared to that of the phonon polaritons (blue solid line) in hBN slabs of corresponding thickness. Both the propagation length and the lifetime of EPhPs increase at frequencies close to the STO (the regime for canalization modes). This can be explained by the large negative real part of $\varepsilon_{eff,x}$ near the STO, which actually reduces the field confinement inside the material, repelling the fields and hence reducing the absorption. We also perform another simulation of placing the dipole to launch EPhPs on the nanograting (metasurface). We fit the propagation length of the dipole-launched canalization EPhP mode with an exponential decay model, which is plotted as an open symbol in the figure. The canalization EPhP on the nanograting exhibits a smaller propagation length (about 780 nm at 1495 cm$^{-1}$) compared to that of EPhPs simulated based on the effective medium model (about 1 μm at 1495 cm$^{-1}$). In the experiment, we observe that the canalization EPhP propagates only 220 nm (solid symbol), which is much smaller than the two theoretical values. This can be explained by higher damping in the experiment caused by fabrication uncertainties and material damage from etching. We also perform the simulations for the cases without the substrate and observe that the polariton propagation length and lifetimes can be largely increased. Therefore, fabricating higher-quality metasurface on a lossless substrate can improve the propagation length of canalization EPhPs.

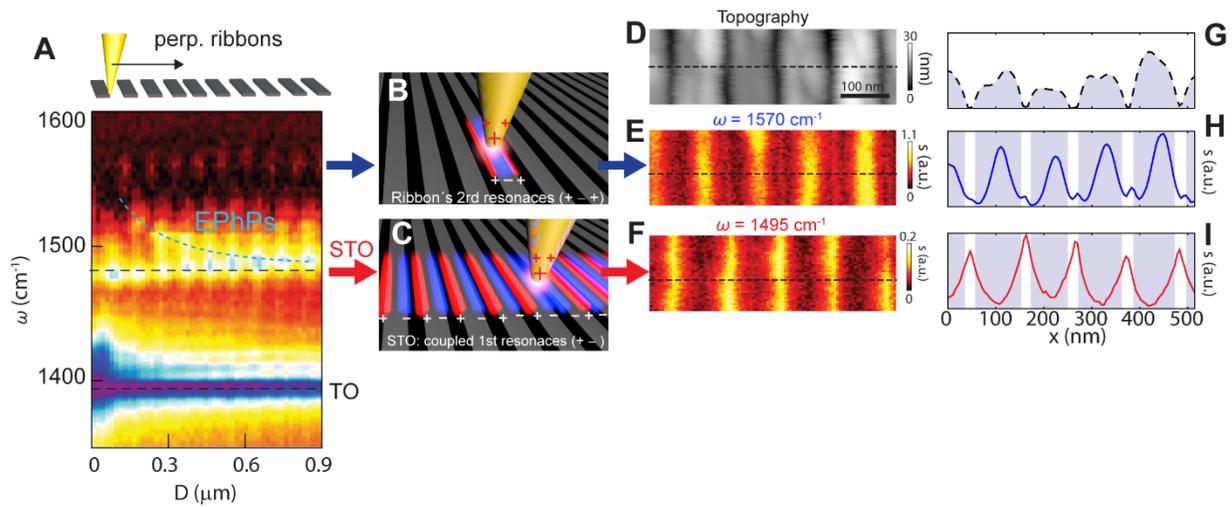

**Figure S7. Near-field imaging of the grating at the fundamental STO (coupled dipolar) resonance and at the second order (+ - +) resonance.** (**A**) Spectroscopic near-field line scan perpendicular to the ribbons (shown in Fig.3B of the main text). (**B** and **C**) schematics of tip excitation of the STO resonance and the ribbons´ 2nd resonance (+ - +), respectively. (**D**) Topography of the grating. (**G**) Line profiles along the dashed line in (**D**). (**E** and **F**) Near-field images measured at the STO resonance ($\omega$ =1495 cm⁻¹) and the ribbons´ 2nd (+ - +) resonance ($\omega$ =1570 cm⁻¹). (**H** and **I**) Line profiles (averaged by 10 lines) along the dashed lines in (**E**) and (**F**). Clearly, the strong near-field signals are observed at the gaps for the STO resonance, while the strong near-field signals are found on the ribbons for the 2nd resonance (+ - +).

Antenna-launched EPhPs  Merged image  FT of (b)

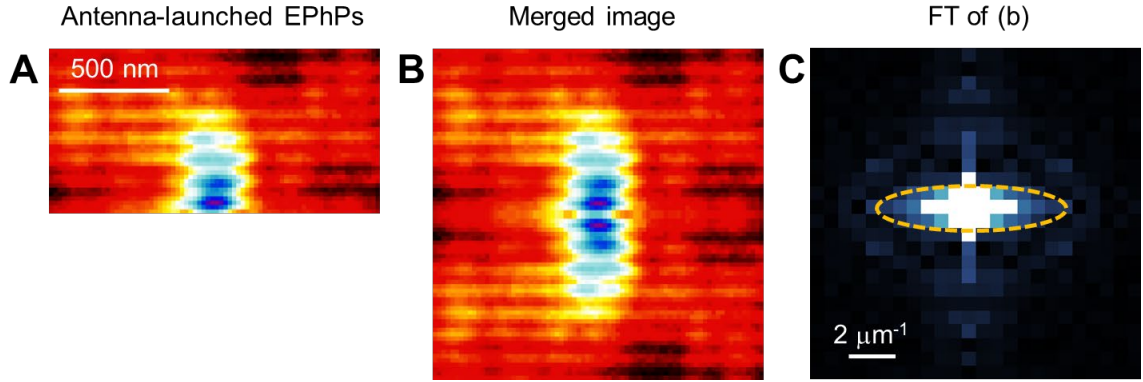

**Figure S8. Determination of experimental polariton isofrequency contours.** (**A**) Near-field distribution of antenna-excited canalization EPhPs (at $\omega= 1495$ cm$^{-1}$, as shown in Fig.4B of main text). (**B**) Image obtained by merging original and vertically flipped image of panel (a). This procedure increases the number of image pixels for the subsequent Fourier transform, in order to increase the number of pixels in the Fourier transform and thus its quality. (**C**) Fourier transforms (FT) of the image shown in panel (B) using the Gwyddion. In this image, we observe a compressed solid ellipse rather than the isofrequency contour (or surface) as theoretically predicted in Fig.1G, which could be because of the low resolution of the FT image. This compressed ellipse can be qualitatively described by the simulated isofrequency contour (dashed orange curve), revealing the canalization features of polaritons observed in real space. The discrepancy between the simulation and the experiment could be due to the size uncertainties of fabricated structures.

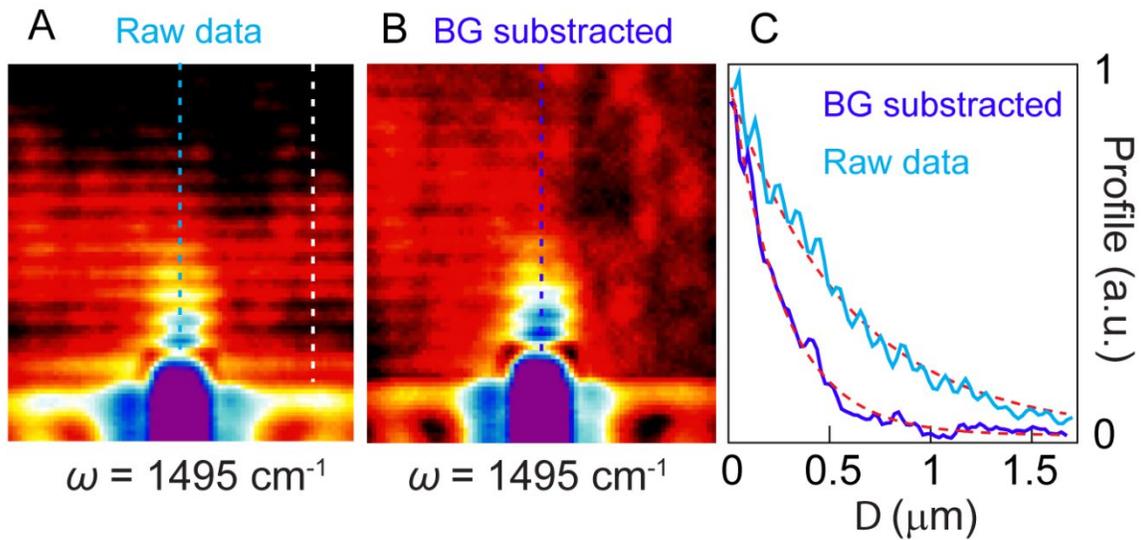

**A** Raw data  **B** BG substracted  **C**

$\omega = 1495 \text{ cm}^{-1}$   $\omega = 1495 \text{ cm}^{-1}$

BG substracted
Raw data

Profile (a.u.)

D (μm)

**Figure S9. Background（BG）subtraction for evaluating the decay length of the antenna-launched canalization PhPs.** (**A**) Near-field image measured at $\omega$ =1495 cm$^{-1}$, as already shown in Fig.4B of the main text. (**B**) The resulting image of (**A**) after subtracting the background (averaged by 10 lines) along the white dashed line marked in (**A**). We performed the background subtraction by using a predefined functionality of an open-source software (Gwyddion). (**C**) Solid lines, profiles along the dashed blue lines in (**A**) and (**B**). Dashed red lines, the fitting of the solid lines. For line profile of the raw data (**A**), we obtain the decay length about 590 nm. After the background subtraction, we obtain the decay length about 220 nm for profile (**B**).